\documentclass[acus]{JAC2000}

\usepackage{graphicx}

\begin{document}
\title{Experimental Determination of the Characteristics of a Positron
Source Using Channeling\thanks{Research made in the framework of INTAS
Contract 97-562}.}

\author{
R.Chehab, R.Cizeron, C.Sylvia (LAL-IN2P3, Orsay, France)\\
V.Baier, K.Beloborodov, A.Bukin, S.Burdin, T.Dimova, A.Drozdetsky, V.Druzhinin,
M.Dubrovin\thanks{presently at Wayne State University at Cornell, Ithaca
NY, USA}, \\ V.Golubev, S.Serednyakov, V.Shary,
V.Strakhovenko  (BINP, Novosibirsk, Russia)\\ 
X.Artru, M.Chevallier, D.Dauvergne, R.Kirsch, Ph.Lautesse, J-C.Poizat,\\
J.Remillieux  (IPNL-IN2P3, Villeurbanne, France)\\
A.Jejcic  (LMD-Universite, Paris, France)\\
P.Keppler, J.Major  (Max-Planck Institute, Stuttgart, Germany)\\
L.Gatignon (CERN, Geneva, Switzerland)\\
G.Bochek, V.Kulibaba, N.Maslov (KIPT, Kharkov, Ukraine)\\	       
A.Bogdanov, A.Potylitsin, I.Vnukov (NPI-TPU, Tomsk, Russia)
}

\maketitle

\begin{abstract} 
Numerical simulations and `proof of principle' experiments showed clearly
the interest of using crystals as photon generators dedicated to intense
positron sources for linear colliders. An experimental investigation, using
a 10 GeV secondary electron beam, of the SPS-CERN, impinging on an axially
oriented thick tungsten crystal, has been prepared and operated between May
and August 2000.

After a short recall on the main features of positron sources using
channeling in oriented crystals, the experimental set-up is described. A
particular emphasis is put on the positron detector made of a drift chamber,
partially immersed in a magnetic field. The enhancement in photon and positron
production in the aligned crystal have been observed in the energy range 5 to
40 GeV, for the incident electrons, in crystals of 4 and 8 mm as in an hybrid
target. The first results concerning this experiment are presented hereafter.
\end{abstract}

\section{Introduction}

   The enhancement of radiation observed, in channeling conditions, in a
crystal with respect to bremsstrahlung, makes crystal targets interesting
for obtaining large positron yields: the high rate of photons generated along a
crystal axis produces a corresponding high positron yield in the same crystal
target \cite{b1}. Association of crystals, where intense radiation takes
place and amorphous targets, where photons are materialized in e+e- pairs,
have also been investigated \cite{b2}. The insertion of this kind of target in
a typical scheme for a positron facility of a linear collider has also been
considered \cite{b3}. The high current of the incident electron beam
on the target leading to important energy deposition in the target and to
possible crystal 
damages, mainly caused by Coulomb scattering of the electron beam on the
nuclei, these two aspects have been studied.
  Simulations of warm crystals, heated by the energy deposited, were
provided in actual working conditions using the JLC (Japanese Linear
Collider) conditions \cite{b3}. Concerning the radiation damages, a beam test has
been performed with the SLC beam; fluences as high as $2\cdot10^{18} mm^{-2}$,
corresponding to hundred hours of continuous working of a linear collider as
JLC, appeared harmless \cite{b3.1}.
  Proof of principle experiments (1992-93, Orsay) \cite{b4} and
  (1996, Tokyo) \cite{b5}
showed photon and positron yields enhancements. An experimental verification
of the yield, energy spectrum and transverse emittance of a crystal positron
source should give definite answers on the useable positron yield for a LC
(Linear Collider)\cite{b6}.
A beam test on the transfer lines of the SPS-CERN started
this spring. After a description of the experimental set-up, with some
emphasis on the positron detector, the first results are provided.

\section{The experimental set-up}

\begin{figure}[t]
\centering
\includegraphics[width=80mm]{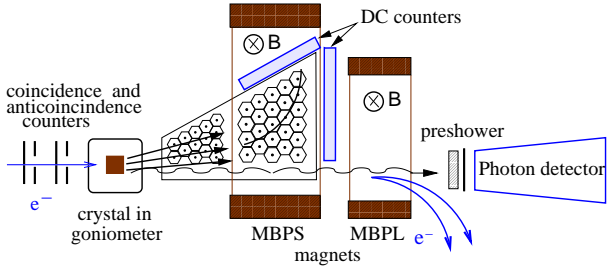}
\caption{Experimental set-up}
\label{exp}
\end{figure}

The experiment is using multi-GeV electron beams of the SPS. The electrons,
after passing through profile monitors and counters (trigger) impinge on the
targets with energies from 5 to 40 GeV (mainly at 10 GeV). Photons, as well
as e+e- pairs are produced in the target. These particles come mainly in the 
forward
direction and travel across the magnetic spectrometer, consisting
of the drift chamber and positron counters inserted between the poles of a
spectrometer magnet (MBPS). The most energetic photons and charged particles
come out nearly in the forward direction. The charged ones are swept by
a second magnet (MBPL) after which the photons reach the photon detector
made of preshowers and a calorimeter (see figure~\ref{exp}).

\section{The main elements of the set-up}
\textbf{The beam:}
the SPS bursts are made of 3.2 seconds duration pulses with a
      14.4 seconds period. $10^4$ electrons/burst are usually obtained at 10
GeV electron energy.

\textbf{The trigger:} 
      the channeling condition requires
      that the incident electron direction angle, with respect to the
      crystal axis, be smaller than the Lindhard critical angle,
      $\Psi_c~=~\sqrt{2\cdot U/E}$
      where $U$ represents the potential well of an atomic row and E the 
      incident electron  
      energy. In order to fulfil it we installed a trigger system made of
scintillator counters.  
      For 10 GeV and $<\!111\!>$ axis for the tungsten crystal, $\Psi_c=0.45
mrad$. 
      Taking into account the presence of crystal effects at angles slightly
      larger than the critical angle, the acceptance angle for the trigger
      has been chosen at $0.75 mrad$. That gives typically a rate of "good
      events" in the experimental conditions of 1\%. 

\textbf{The target:}
two kinds of tungsten targets have been installed on a
      0.001 degree  precision goniometer: a 8 mm thick tungsten crystal
      and an hybrid target made of 4 mm crystal and 4 mm amorphous.
      Mosaic spreads of both crystals are less than $0.5 mrad$.

\textbf{The positron detector:}
the drift chamber  is made of hexagonal cells, filled with
	   a gas mixture $He(90\%) CH_4(10\%)$ and presents two parts:\\
\textbf{the first part (DC1)}, with a cell radius of 0.9 cm, is escaping
           mainly the magnetic field of the bending magnet MBPS. It allows
	   the measurement of the position and exit angle of the emitted
           pairs.\\ 
\textbf{the second part (DC2)}, with a cell radius of 1.6 cm, is
	   submitted to the magnetic field. It allows the measurement of
           the positron momentum. Two values of the magnetic field are used,
	   1 and 4 kGauss, to investigate the two momenta regions:
	   from 5  to $20 MeV/c$ and from 20 to $80 MeV/c$.

Signal and field wires are short (6 cm) and made of gold plated
	tungsten and titanium, respectively. Counters (scintillators) are
        put on the lateral walls in order to define the useful region.
	The drift chamber exhibits 21 layers and the resolution is of 300
	microns. The maximum horizontal angle being accepted is 30 degrees.
        The limited vertical size sets the overall acceptance of the chamber
	to 6\%. The choice of Helium provides a small multiple scattering
	($0.001 X_0$).
	
\textbf{The electronics:}
for drift time measurements the electronics
	    detects the leading edge of the signal, coming from the sense
            wire, and digitize the time with 3ns resolution. Front end
            electronics is made of preamplifier, shaper and ECL discriminator.
	    The TDC (Time to Digital Converter) have a scale range of
	    1.5 microseconds. A common stop is used.
	    
\textbf{The photon detector:}
            Photon multiplicity is rather high: about 200 photons/event for
	    a 8 mm thick tungsten crystal oriented along its $<\!111\!>$ axis.
	    The angular acceptance of the photon detector is 5.5 mrad (half
	    cone angle). That gives 2000 triggered photons in an SPS burst.
            The photon detector is made of:\\
2 preshowers giving the relative photon multiplicity (one with nominal 5.5
mrad acceptance and the other with reduced 1.5 mrad acceptance),\\
a NaI calorimeter providing the energy radiated.
	     The role of the photon detector is essential for operating the 
	     crystal orientation.
	     
\section{First results}
\begin{figure}[htb]
\centering
\includegraphics*[width=60mm]{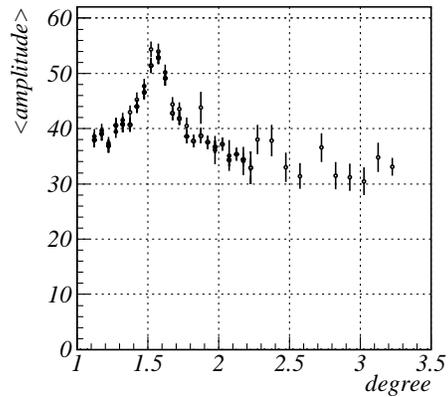}
\caption{Rocking curve measured on preshower for the hybrid target.}
\label{rock}
\end{figure}
\textbf{Enhancement in photon production.}
   On the $<\!111\!>$ axis of the tungsten crystals, the ultrarelativistic
   electrons
   radiate more photons than by classical bremsstrahlung. The preshower
   provides the relative photon multiplicity with respect to the crystal
   orientation. We reported, on figure~\ref{rock}, the associated rocking
   curve
\begin{figure}[htb]
\centering
\includegraphics*[width=60mm]{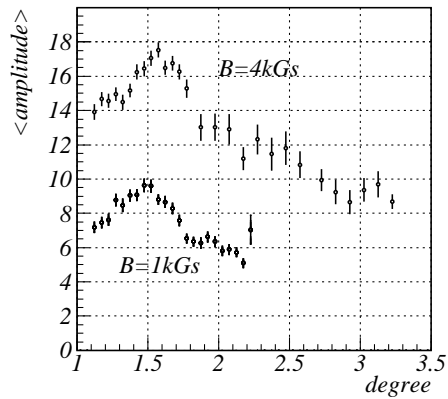}
\caption{Rocking curves measured on DC counter 1 for the hybrid target with
two values of the magnetic field.}
\label{DCcounters}
\end{figure}
   for the 8 mm hybrid target, for an incident energy of 10 GeV. 
   The enhancement is about~1.8.

\begin{figure}[t]
\centering
\includegraphics*[width=60mm]{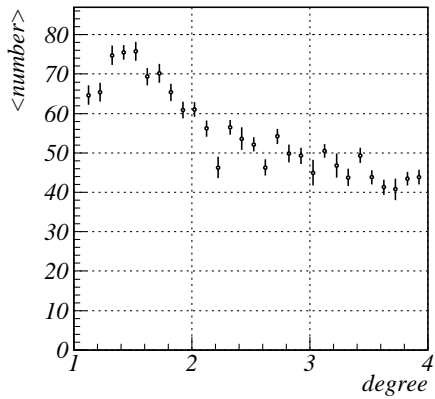}
\caption{Rocking curve measured with number of wires in DC for the hybrid
target.}
\label{DCwires}
\end{figure}
\textbf{Enhancement in positron production.}
   A corresponding enhancement in positron production is observed in
   channeling conditions as can be seen on figure~\ref{DCcounters}. Rocking curves for the
   positrons use the informations from the two counters put on the lateral
   walls of the Drift Chamber. Counter 1 is receiving low energy positrons
   strongly deflected by the magnetic field (1 and 4 kGauss) and counter 2,
   the high energy positrons (and the electrons) weakly deflected. It can be
   seen that the ratio aligned/random is about 2, for an incident energy of
   10 GeV, in agreement with the simulations (Counter 1). The enhancement 
   in positron production has also been observed with the number of hitted
   wires in the Drift Chamber (see figure~\ref{DCwires}).
\begin{figure}[htb]
\includegraphics*[width=85mm]{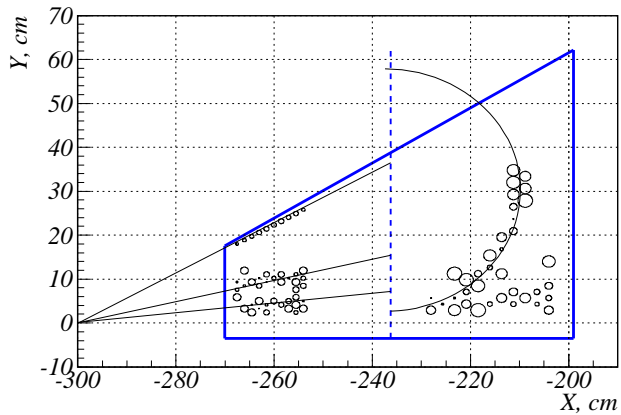}
\caption{Typical event for the hybrid target. Field $0.4T$}
\label{event}
\end{figure}

\textbf{Positron tracks.}
   Positron tracks, for different running conditions (Beam energy, crystal
   thickness, magnetic field) have been reconstructed. We present on
   figure~\ref{event}
   an example of these tracks reconstructed separately in the two parts of
   the chamber on the basis of the signals
   delivered by the Drift Chamber.
   
\begin{figure}[htb]
\centering
\includegraphics*[width=60mm]{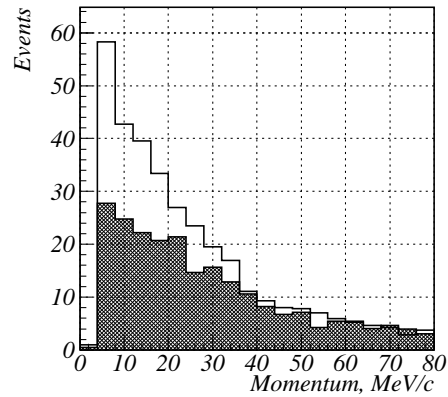}
\caption{Positron momentum distribution measured by DC with hybrid 
target. Field 1 kGauss. Incident energy is 10 GeV. The unfilled histogram
represent the oriented crystal, dark one --- the target in random position.}
\label{momentum}
\end{figure}
\textbf{\noindent Momentum distribution.}
   Preliminary momentum distribution has been determined for the accepted
positrons in the Drift Chamber (see for example figure~\ref{momentum}).
Such distribution has also been obtained for the two values of
the magnetic field: 1 and 4 kGauss for several energies (from 5 GeV to 40
GeV).

\textbf{Incident energy dependence.}
   Variation of the incident energy from 5 to 40 GeV, clearly shows
   enhancement in photon as in positron production with growing energy;
   At 40 GeV, the ratio axis/random is more than 4.

\section{Summary and Conclusions}
   These first results on positron production in channeling conditions with
   high energy electrons clearly show an enhancement in photon as in
   positron production in agreement with the simulations. 
   For the positrons the enhancement takes place mainly at low energy.
   This aspect is of
   particular importance for the positron capture in the accelerator channel.
   The Drift Chamber operates properly to determine the positron track and,
   hence to provide the exit angle and positron momentum.
   These first results, which give a clear confirmation for the interest of
   using this kind of sources will be completed by an appropriated analysis
   in the next months.

\end{document}